# Chiral Landau levels in Weyl semimetal NbAs with multiple topological carriers

Xiang Yuan [†,1,2], Zhongbo Yan [†,3], Chaoyu Song [†,1,2], Mengyao Zhang[5,6], Zhilin Li[4,6], Cheng Zhang[1,2], Yanwen Liu[1,2], Weiyi Wang[1,2], Minhao Zhao[1,2], Zehao Lin[1,2], Tian Xie[1,2], Jonathan Ludwig[7], Yuxuan Jiang[7], Xiaoxing Zhang[8], Cui Shang[8], Zefang Ye[1,2], Jiaxiang Wang[1,2], Feng Chen[1,2], Zhengcai Xia[8], Dmitry Smirnov[7], Xiaolong Chen[4,6], Zhong Wang[3,4], Hugen Yan[1,2*], Faxian Xiu[1,2,9*]

[1]State Key Laboratory of Surface Physics and Department of Physics, Fudan University, Shanghai 200433, China

[2]Collaborative Innovation Center of Advanced Microstructures, Fudan University, Shanghai 200433, China

[3]Institute for Advanced Study, Tsinghua University, Beijing 100084, China

[4]Collaborative Innovation Center of Quantum Matter, Beijing 100871, China

[5]International Center for Quantum Materials, School of Physics, Peking University, Beijing 100871, China

[6]Beijing National Laboratory for Condensed Matter Physics, Institute of Physics, Chinese Academy of Sciences, Beijing 100190, China

[7]National High Magnetic Field Laboratory, Tallahassee, Florida 32310, USA

[8]Wuhan National High Magnetic Field Center, Huazhong University of Science and Technology, Wuhan 430074, China

[9]Institute for Nanoelectronic Devices and Quantum Computing, Fudan University, Shanghai 200433, China

[†]These authors contributed equally to this work.

[*]Correspondence and requests for materials should be addressed to F. X. (E-mail: Faxian@fudan.edu.cn) and H. Y. (E-mail: Hgyan@fudan.edu.cn)

**Abstract**

Recently, Weyl semimetals have been experimentally discovered in both inversion-symmetry-breaking and time-reversal-symmetry-breaking crystals. The non-trivial topology in Weyl semimetals can manifest itself with exotic phenomena which have been extensively investigated by photoemission and transport measurements. Despite the numerous experimental efforts on Fermi arcs and chiral anomaly, the existence of unconventional zeroth Landau levels, as a unique hallmark of Weyl fermions which is highly related to chiral anomaly, remains elusive owing to the stringent experimental requirements. Here, we report the magneto-optical study of Landau quantization in Weyl semimetal NbAs. High magnetic fields drive the system towards the quantum limit which leads to the observation of zeroth chiral Landau levels in two inequivalent Weyl nodes. As compared to other Landau levels, the zeroth chiral Landau level exhibits a distinct linear dispersion in z momentum direction and allows the optical transitions without the limitation of zero z momentum or $\sqrt{B}$ magnetic field evolution. The magnetic field dependence of the zeroth Landau levels further verifies the predicted particle-hole asymmetry of the Weyl cones. Meanwhile, the optical transitions from the normal Landau levels exhibit the coexistence of multiple carriers including an unexpected massive Dirac fermion, pointing to a more complex topological nature in inversion-symmetry-breaking Weyl semimetals. Our results provide insights into the Landau quantization of Weyl fermions and demonstrate an effective tool for studying complex topological systems.

**Introduction**

Dirac semimetals can be viewed as special Weyl semimetals with spin degeneracy.[1] The lifting of the Weyl nodes degeneracy can be accessed by breaking inversion symmetry (IS) in Dirac semimetals. It was recently realized in non-centrosymmetric compounds NbAs family.[1, 2, 3, 4, 5, 6, 7, 8, 9, 10, 11, 12, 13] Such a Weyl semimetal phase is featured by Fermi arcs and chiral anomaly, both of which were intensively investigated.[6-16] The Fermi arcs with unique spin texture have been observed by photoemission on the surfaces of Weyl semimetals.[7, 9, 10, 12, 13, 14] The open-contour feature distinguishes themselves from traditional surface states and results in novel cyclotron orbits under magnetic fields.[15] Meanwhile, the predicted chiral or Adler-Bell-Jackiw anomaly, a chirality imbalance phenomenon in Weyl semimetals, has been widely observed[16, 17, 18, 19, 20, 21, 22, 23, 24] although the origin of some experiments is still under debate.[25, 26] Besides the Fermi arcs and chiral anomaly, one of the most important features of Weyl semimetals is the presence of the zeroth chiral Landau level which is linearly dispersed in $k_z$ direction.[27, 28, 29] It was theoretically predicted to be a hallmark of Weyl fermions that leads to the widely-observed chiral anomaly in Weyl semimetals.[16, 17, 18, 19, 20, 21, 22, 24, 30] However, directly detecting this distinct zeroth Landau level through conventional photoemission or transport experiments remains an unprecedented challenge since both magnetic field and Landau level spectroscopy are required.[27] Also, experimentally approaching the quantum limit is strictly demanded to distinguish the chiral Landau level from other degenerated Landau levels.

Here, we investigate the quasi-particle dynamics in Weyl semimetal NbAs by means of magneto-optics. The ability to detect the Landau levels via the optical reflectance under high magnetic fields enables the observation of the zeroth chiral Landau levels and complex topological nature near the quantum limit. Two sets of inequivalent Weyl cones with different Fermi velocity were identified through $\sqrt{B}$ and $\sqrt{n}$ dependence of the optical transition frequency. The intra-band transition edges were observed as a direct consequence of the chiral Landau levels in both of inequivalent Weyl nodes. The dramatic distinction from other Landau levels, however, is the linear dispersion in $k_z$ direction, resulting in a unique field evolution of transition energy. Detailed study of the transitions between the chiral and normal Landau levels reveals a large particle-hole asymmetry in Weyl cones with higher Fermi velocity which agrees well with the predictions from first-principle calculations[31]. Finite scattering helps to enhance the feature from transition edge.[27] Also, corroborated with our theory, five kinds of optical transitions involving the chiral and conventional Landau levels are unambiguously distinguished which construct the whole band structure of NbAs under magnetic field as summarized in the end. Besides these two inequivalent Weyl nodes, we also reveal the coexistence of massive Dirac fermions as well as trivial band insulating state, suggesting a more complex topological nature in the Weyl semimetal family. The comparison among different optical transitions in NbAs exhibits the uniqueness of the chiral Landau levels. In the following, we will describe these five optical transitions and distinguish their origins one by one, which show the clear contribution from the chiral Landau levels. The comprehensive study of Landau level spectroscopy in NbAs provides insights into multiple novel topological carriers and constitutes an ideal experimental tool for studying complex quasi-particles with their own topological nature.

**Results**

**Single crystal NbAs**

NbAs crystalizes in body-centered tetragonal unit cell with space group $I4_1md$ and point group $C_{4V}$,[32] as shown in Fig. 1a. Since this type of structure lacks a horizontal mirror plane, the inversion symmetry in NbAs is naturally broken which leads to the lift of the spin degeneracy, necessary for developing the Weyl semimetal phase. *ab initio* calculations predict 24 Weyl nodes in the first Brillouin zone of NbAs,[1, 7] where they always come in pairs with opposite chiralities or monopole charges. The studied single crystals were grown by transport method[33] and their crystalline orientation and chemical ratio were examined by X-ray diffraction (XRD) and electron-dispersive x-ray spectroscopy (EDX), respectively (Fig. 1b and Supplementary Figure 1). Three prominent peaks can be clearly identified as (004), (008) and (0012) crystal planes.[32, 34] The lattice constant is calculated to be $a$=3.45Å and $c$=11.68Å which is consistent with the previous studies.[32] The stoichiometry of the sample confirmed by the EDX ensures the comparatively low Fermi level and less undesired defects.[35]

**Magneto-optical spectrum**

Magneto-infrared measurements have been proved to be an effective tool for study Landau quantization for topological materials which can provide rich information such as band parameters and field response.[36] The measurements of magneto-infrared spectroscopy are combined with magneto-transport where the important parameters are extracted that help to understand Landau levels. Here we take magneto-optical spectrum on the (001) crystal surface of NbAs as shown in Fig. 1b inset. The schematic experimental setup of the magneto-optical measurements is displayed in Fig. 1c. Faraday geometry were used in a superconducting magnet with the magnetic field up to 17.5T. More experimental details are available in the method section. As shown in Fig. 1d, the reflection spectra were normalized by the zero-field data. A series of reflection features can be resolved. The reflection maxima and minima systematically shift towards the higher energy with increasing magnetic field. Given the high mobility of NbAs at low temperatures, the field-dependent peaks can be assigned to the Landau level transitions. Although all of the peak energies increase with field, certain differences are discernible. In Fig. 1d and 2a, some of the peak energies increase linearly with the magnetic field (denoted as M type in dark green) while the others do not. The observed difference indicates the distinct origins of low-energy excitations in the system. For a classical system with a quadratic energy dispersion $\varepsilon = \hbar^2 k^2 / 2m^*$, the Landau quantization under magnetic field reads

$$E_n = (n+1/2)\hbar eB/m^*, \qquad (1)$$

where $n, e, \hbar$ represent the Landau index, the elementary charge and the reduced Plank's constant, respectively. Optical selection rules require $\Delta n = |n_i| - |n_f| = \pm 1$, where $n_i, n_f$ denote the Landau level index of the initial and final states. The Landau level transition with the T-type and C-type features are in stark contrast to the classical picture while the M-type peaks fit well to the linear-in-*B* rule (Fig. 2a). The notations C/T/M are named after several terms: Chiral Landau level, normal Transition and Multiple bands which will be discussed in the following sections. All these salient features are reproducible as demonstrated in Supplementary Note II from another NbAs sample of the same batch with an identical growth condition.

**Origin of the inter-Landau-level transitions**

To quantitatively study the spectrum and understand the origin of each transition, we extracted the frequency of inter-Landau-level transitions and plotted it against the magnetic field in Fig. 2a. Generally, Landau level transition induces a peak in the optical conductivity. In order to extrapolate the transition energy, an absolute reflection spectrum with gold overcoating is required (Supplementary Note III). Then the experimental data were fitted to the Drude-Lorentz model (Supplementary Note X) in the presence of magnetic field.[37, 38, 39, 40] The fitted curve in Fig. 2b shows a good consistency with the original experimental data for all magnetic fields (Supplementary Figure 8 in Supplementary Note X). The extracted transition energy typically locates nearby but with a lower energy than where the apparent peak is positioned in the original data (especially for T-type and M-type transitions). The inter-Landau-level transition energy in Fig. 2a is obtained from the aforementioned fitting scheme.

Fan diagram reveals rich features that can be used to determine the electronic structure of NbAs under magnetic fields. A series of transition energies follow a straight line pointing to the origin of the plot on a scale of $\sqrt{B}$ which provides key information for the T-type transitions as shown in Fig. 2a. Massless Weyl fermions are described by the Hamiltonian $H(\mathbf{k}) = v_F \mathbf{k} \cdot \sigma$ ($\hbar = 1$ for notation simplicity) where $\sigma = (\sigma_x, \sigma_y, \sigma_z)$ and $v_F$ represent the Pauli matrices and the Fermi velocity, respectively. With the introduction of a magnetic field along $z$ direction, the formed Landau levels read[18],

$$E_n(k_z) = \begin{cases} \text{sgn}(n) v_F \sqrt{k_z^2 + 2|n|eB}, & |n| \geq 1 \\ -v_F k_z & n = 0. \end{cases} \quad (2)$$

The lowest Landau level is of chiral nature which makes it quite different from the higher Landau levels. Based on Eq. (2), it is readily found that the density of states is given by

$$D(\varepsilon) = \frac{eB}{4\pi^2 v_F} [1 + 2 \sum_{n=1} \text{Re}(\frac{|\varepsilon|}{\sqrt{\varepsilon^2 - 2|n|eBv^2}})]. \quad (3)$$

$\Sigma$ is a sum symbol. From Eq. (3) and Fig. 2c, the singularity occurs at $k_z = 0$ where the density of state becomes infinite. Therefore, the optical transition energy should follow $\sqrt{B}$ and $\sqrt{n}$ law for Weyl semimetals with non-zero index. Experimentally, all the T-type transitions clearly adopt the $\sqrt{B}$ dependence, indicating the Landau level transition from non-zero index (illustrated by red arrows in Fig. 2c) and thus serving as a direct evidence of ultra-relativistic fermions.[29, 41, 42, 43, 44] Note that all the schematic plots of Landau level dispersion are also applicable to the hole-doped Weyl semimetals because the optical transition is schematically symmetric, *i.e.*, it has the same absorption energy for the electron/hole doping with the same carrier density.

We now assign the specific Landau level index *n* and verify the $\sqrt{n}$ dependence. Deduced from Eq. (2) and the optical selection rules, the resonance frequency[43] or the slope of T1-T7 in Fig. 2a, designated as S1-S7, should be proportional to $\sqrt{n+1} \pm \sqrt{n}$, where the positive (negative) sign is for inter-band (intra-band) transition. Apparently, T1-T7 do not simply follow this rule and the observed optical transition cannot be explained by the Landau quantization of a single ideal Weyl cone. In fact, we have noticed that the ratio of the adjacent slopes keeps nearly a constant, *i.e.*, $S7/S6 \approx S6/S5 \approx S5/S4 \approx S3/S2$. This clearly shows signatures of two inequivalent Weyl nodes. By further calculating $S7:S6:S5:S3 \approx (\sqrt{4}+\sqrt{5}):(\sqrt{3}+\sqrt{4}):(\sqrt{2}+\sqrt{3}):(\sqrt{1}+\sqrt{2})$, these four transitions can be identified to originate from one Weyl node, denoted as Weyl node 1 (W1). Then the peak assignments can be made as follows: T3 corresponds to $L_{-1} \rightarrow L_2$ or $L_{-2} \rightarrow L_1$ and T5 $L_{-2} \rightarrow L_3$ or $L_{-3} \rightarrow L_2$, and so on so forth for T6 and T7. Here $L_n$ represents the $n^{th}$ Landau level. Note that only $L_{-n} \rightarrow L_{n+1}$ is presented with labels for simplicity. These assignments can be further verified by the fan diagram in Fig. 2d, where $\sqrt{n+1}+\sqrt{n}$ dependence can be clearly seen. The assignments for the second Weyl node (W2, Fig. 2e) involving T6, T5, T4 and T2 can be made in the same manner as that for W1. Therefore, we elucidate the existence of two inequivalent Weyl nodes and the Landau level energy is indeed proportional to $\sqrt{B}$ and $\sqrt{n}$. By performing a linear fit, the Fermi velocity is derived to be $v_{F1} = 2.0 \times 10^5$ m/s and $v_{F2} = 1.7 \times 10^5$ m/s which matches well with the previous first-principle calculations.[31] Weyl nodes 1 and 2 are illustrated in red and blue in Fig. 2, respectively. It is worth mentioning that T5 and T6 include the transitions from both Weyl nodes and therefore are labelled in purple. This overlapping is a natural consequence of two inequivalent Weyl cones as shown in Supplementary Note X. Usually for the system close to quantum limit, the inter-band transition with higher index should show smaller amplitude. Here the comparatively higher amplitude of T5, T6 can be explained by the overlapping of these close transitions. The larger width of T5, T6 originates from same mechanism. To distinguish more clearly how those different electronic bands influence the optical transitions, Figur 2a is replotted and divided into several panels for further clarification in Supplementary Note IV and Supplementary Figure 4. All the observed features originated from different bands are also summarized in the end.

**Quantum limit for inequivalent Weyl nodes**

The existence of resonant peaks corresponding to $L_{-1} \rightarrow L_2$ or $L_{-2} \rightarrow L_1$ suggests that the system is close to the quantum limit. The Fermi velocity and the energy offset between two inequivalent Weyl nodes are also suggested by angle-resolved photoemission spectroscopy.[9] Different from inter-band transitions which always come in pairs (blue and red for two Weyl nodes), as shown in Fig. 2a in the low frequency regime, there exists only one linear-in-$\sqrt{B}$ transition (T1) which leads us to check the position of Fermi level and the quantum limit. Therefore, magneto-transport measurements were carried out with a Hall bar device on the same sample with the same geometry as the magneto-optical measurements. Fig. 3a and 3b present the longitudinal MR in different field regimes. Clear Shubnikov-de Haas (SdH) oscillations can be resolved on a large upward background. The ultimate quantum limit is reached near a field of 18T similar to the previous magnetic torque measurements.[15] A large MR reaches 60000% which features the compensated material and topological massless fermion with the suppressed backscattering.[45] We extracted the peak positions of the resistivity and assigned the integer index. Two sets of SdH oscillations were found as shown in Fig. 3c (also, arrows in different colors in Fig. 3a and b). It is noted that the closed cyclotron orbit under magnetic field follows the Lifshitz-Onsager quantization rule $S_F \frac{\hbar}{eB} = 2\pi(n + \frac{1-\phi_B}{2})$, where $S_F, \phi_B$ are Fermi surface area and Berry phase, respectively. Performing the linear fit in the fan diagram yields the intercepts which are nearly zero, indicative of the topological non-trivial Berry phase. The non-trivial topology of the oscillation components evidences two gapless Weyl nodes. Consistent with the previous reports,[46, 47] our Hall effect measurements suggest a typical two-carrier transport behavior (Supplementary Note V and Supplementary Figure 5) and the quantum limit is reached with magnetic field 19T and 12T for the inequivalent Weyl nodes. The fact that one of the intra-band transitions is missing can now be explained by the entrance of the quantum limit.

**Chiral Landau levels**
It has a profound influence on intra-band transitions when the system reaches the quantum limit. As shown in Fig. 3d, different from the discussed T-type or M-type transitions, there remains three other peaks denoted as C1 to C3 which follow neither linear-in-$B$ nor linear-in-$\sqrt{B}$. Since all the allowed transitions with non-zero index have been exclusively identified at high energy, the remaining C1 to C3 are anticipated to come from the intra-band transitions between chiral $L_0$ and $L_{\pm 1}$. The linearly-dispersed zeroth Landau level in Weyl semimetal is unique compared with traditional systems. It has been used to explain the four-fold splitting in magneto-optical spectrum of ZrTe$_5$ and was further studied by circular polarization resolved experiments.[43, 48] It also differs from other topological materials such as Kane semimetal HgCdTe[49] or Dirac semimetal graphene[50] where a flat zeroth Landau level along $k_z$ locates at zero energy regardless of various magnetic fields, resulting in an energy gap between the zeroth and the first Landau level. A previous magneto-optical study also suggests $Cd_3As_2$ to be Kane semimetal partly due to the lack of chiral Landau level.[29] The transitions involving the zeroth Landau level should occur at $k_z = 0$ in Kane or 2D Dirac semimetals while a fixed zeroth Landau level should result in a linear-in-$\sqrt{B}$ relation for $L_0 \to L_1$.[44, 49, 50, 51] As to Weyl semimetals, the 3D structure with Weyl cones ensures a different form of zeroth Landau level, as described by Eq. (2), with unique optical features that agree well with the observed C-type transitions in the following four aspects. Firstly, for the lowest energy intra-band transition, it does not necessarily occur at $k_z = 0$ due to the lack of density of state maxima. Instead, the allowed transition has a certain frequency edge where the density of states experiences a sudden increase from zero when the Fermi level crosses the zeroth Landau level as shown in Fig. 3e and 3f (colored arrows). Another fact that needs to point out is that in the presence of impurity scattering, the edge for the transition involving chiral Landau level will be enhanced and the peaks for inter-band transitions will be suppressed[27] (see details in Supplementary Note). Thus, it is reasonable to expect that the peak due to the transition edge has a similar feature with comparable amplitude as the feature from inter-band transitions. Secondly, the energy of the normal Landau level transitions, for example T1-T7, is determinedly by the distance between Landau levels at $k_z = 0$ which is proportional to $\sqrt{B}$. Varying the Fermi level can only allow or abandon certain optical transitions, but it cannot affect the transition energy. For the linearly-dispersed chiral Landau level, however, the transition edge energy is determined by the Fermi level as well, resulting in non-$\sqrt{B}$ field evolution; Thirdly, to observe the optical feature of the zeroth Landau level transition, it requires the system near the quantum limit (*i.e.*, low-carrier-density sample under high magnetic fields). For the system far from that, the transition energy rapidly decreases to a vanishing value beyond our experimental resolution (Supplementary Note VI and Supplementary Figure 6, the experimental challenge of probing chiral Landau level); Fourthly, the optical behavior involving the zeroth Landau levels can be significantly influenced as the system approaches the quantum limit. Here, in our field regime, one of the Weyl nodes does not approach the quantum limit. Therefore, the Landau level intersects with the first Fermi level resulting in three intra-band transition edges as displayed in Fig. 3e. The upper two have the exactly same energy (Supplementary Note VIII). For those two transitions at same $k_z$, the lower one is the transition between chiral Landau level and $1^{st}$ Landau level while the upper one is the transition between $1^{st}$ and $2^{nd}$ Landau level. Further increasing the magnetic field enlarges the transition energy (Fig. 3d, red line). As long as the system reaches the quantum limit (Fig. 3f), the right edges will disappear since the intersection between Fermi level and $L_1$ no longer exists. The behavior of C3 transition fits well to the discussed scenario and the peak disappears near 12 T.

Next, we performed a quantitative study for these intra-band transitions. Previously, the first-principle calculations reveal that NbAs is particle-hole asymmetric.[31] To capture the effect of particle-hole asymmetry in the massless system, we can add a diagonal term to the low-energy Hamiltonian that describes the Weyl fermions with different Fermi velocity on conduction and valence band[31] by

$$H_\alpha(\mathbf{k}) = \lambda_\alpha \sqrt{\eta_1^2 k_x^2 + \eta_2^2 k_y^2 + \eta_3^2 k_z^2} + v_\alpha \mathbf{k} \cdot \sigma \qquad (4)$$

where $\alpha=1,2$ corresponds to the Weyl node reaching or not reaching the quantum limit, and $\lambda_\alpha$ denotes the effect of asymmetry. For $\lambda_\alpha \neq 0$, the conduction and valence bands are asymmetric. For simplicity, we assume $\eta_1 = \eta_2 = 1$ and $\eta_3 = 0$. Similar to the previous derivation of Landau level, we consider that a magnetic field is exerted in $z$ direction, then the Landau level reads

$$E_{n,\alpha}(k_z) = \begin{cases} \lambda_\alpha f_1(n,B) + \text{sgn}(n) v_\alpha \sqrt{(k_z - f_2(n,B))^2 + 2eB|n|}, & |n| \geq 1 \\ \lambda \sqrt{eB} - v_\alpha k_z, & n = 0 \end{cases} \qquad (5)$$

$f_1(n,B) = (\sqrt{(2n+1)eB} + \sqrt{(2n-1)eB})/2, \quad f_2(n,B) = \lambda_\alpha(\sqrt{(2n+1)eB} - \sqrt{(2n-1)eB})/2v_\alpha$

As shown in Supplementary Note VII, the frequency influenced by the asymmetry is always orders of magnitude lower for the inter-band transition frequency. However, the effect on the intra-band transition cannot be neglected. We define the energy of the low (the left arrow in Fig.3e and the arrow in Fig. 3f) and high (the bottom-right arrow in Fig. 3e) transition edges as $\omega_{0\to 1,1}$ and $\omega_{0\to 1,2}$. For $\omega_{0\to 1,1}$, the Fermi level intersects with chiral Landau level, therefore $\omega_{0\to 1,1} = E_{1,\alpha} - |E_F|\big\|_{v_\alpha k_z = \lambda_\alpha \sqrt{eB} - |E_F|}$ can be calculated. For the high energy edge $\omega_{0\to 1,2}$ from the Weyl node not in the quantum limit, the Fermi level intersects with the first Landau level. In the same manner, $k_z$ is firstly determined and then $\omega_{0\to 1,2} = |E_F| - E_{0,\alpha}$ can be calculated. Using the similar approach for the first and second Landau levels, we can calculate $\omega_{1-2}$ for T1 as well. The energy of the transition edge becomes a function of $E_F, v, \lambda, B$. All details including the derivation of equations and data fitting are available in Supplementary Note VII. For a given band parameter, the position of Fermi level determines the quantum limit. Therefore, $E_F$ is no longer a free parameter for the data fitting. C2 transition can be best fitted only by higher energy edge $\omega_{0\rightarrow1,2}$ as shown in Fig. 2a and 3d (fitting curves plotted as red solid line) which means that C2 is originated from the chiral Landau level in W1 (not in quantum limit). The fitting parameters are $v_1' = 2.2\times10^5$ m/s and $\lambda_1 = 1.1\times10^5$ m/s which agree well with the previously-discussed origin of W1. When putting $v_1', \lambda_1$ into $\omega_{0\rightarrow1,1}$ and attempting to extract the position of the lower edge from W1, the calculated curve is almost overlapping with T1 as shown in Fig. 2a and 3d (red solid line).

For C1 from W2, the best fit for $\omega_{0\rightarrow1,1}$ gives $v_2' = 1.7\times10^5$ m/s and $\lambda_2 = 0.26\times10^5$ m/s. The fitted curve is plotted as lower blue solid line in Fig. 3d. By examining the slope of T1 and $\omega_{1-2}$, the particle-hole asymmetry parameter is determined to be $\lambda_1' = 1.0\times10^5$ m/s. Note that the equations in tables and figures use the Weyl model without particle-hole asymmetry for simplicity (see more details in Supplementary Note VIII). Using the fitted parameters from W2, one can predict the positions of higher frequency edge $\omega_{0\rightarrow1,2}$ from W2 as shown by the upper blue lines in Fig. 3d. The experimentally extracted C3 peaks fit well to the predicted curve. Therefore, the C3 peak comes from the higher frequency edge transition $\omega_{0\rightarrow1,2}$ from W2. In both Weyl nodes, the frequency of higher edge increases faster with magnetic field than the lower edge. As also determined by the transport measurements, W2 reaches the quantum limit near 12 T. Since the intersect between the Fermi level and the first Landau level is gone, it is reasonable to find that C3 peak disappears near 12 T.

The fitting parameters $v, \lambda$ further verify the anticipated origin of the intra-band transitions accompanied by the following independent evidences: (i) Experimental data points from all chiral Landau level transitions fit well with the magnetic field evolution as predicted by theory[27, 29] that clearly differs from the contribution of other bands. (ii) The extracted velocity $v$ from C1 and C2 is very close to the value given by inter-band transitions where $v$ is acquired by $\sqrt{B}$ and $\sqrt{n}$ fitting. The edge behavior agrees well with our analysis for different situations (reaching the quantum limit or not). (iii) The extracted particle-hole asymmetry generally agrees with the features predicted by previous density functional theory (DFT) calculations,[31] where the Weyl node with higher Fermi velocity possesses a much stronger particle hole asymmetry. Our conclusion of $\lambda_1 > \lambda_2$ accords well with this prediction. The difference of Fermi velocity between particle and hole bands in previous DFT is 92% and 6% for W1 and W2,[31] respectively, while as a comparison our Landau level spectroscopy gives 53% and 15%. With such a quantitative analysis, we indeed verify the existence of the chiral Landau level and particle-hole asymmetry in NbAs. To quantitatively describe all 10 features (C1-C3 and T1-T7) in the spectra, it requires only two parameters of $v$ and $\lambda$ with a simple Weyl particle picture as described by Eq. (4).

**Discussion**

Other than C/T-type transition, there exist several M-type transitions at higher energy (also see the mid-infrared spectra in Supplementary Note IX and Supplementary Figure 7) which follow linear-in-$B$ evolution and clearly possess positive intercepts as shown in Fig. 4a. These features usually indicate a gapped structure corresponding to massive fermions[52, 53] which could coexist with Weyl band. Distinct from M4-M5 transitions in Fig. 2a and 4a where the slope ratio between the adjacent ones is 2:1 (reasonable in classical quadratic band). However, a very large slope ratio is found between M1 and M2. To understand this, we have provided the main physics of an inversion-symmetry-breaking Weyl semimetal by a four-band minimal model whose form takes

$$H(k) = v_x k_x \tau_x \sigma_z + v_y k_y \tau_y + v_z k_z \tau_x \sigma_x + (m - Ck^2)\tau_z + \Delta \tau_y \sigma_x \qquad (6)$$

where $v_{x,y,z}$ represent the velocities, $m$ and $C$ are two constants, and $\Delta$ denotes the inversion-symmetry-breaking term. With the introduction of magnetic field in $z$ direction, the Landau levels corresponding to the two gapped bands follow

$$E_n(k_z) = \begin{cases} \text{sgn}(n)\sqrt{v^2 k_z^2 + (\sqrt{2v^2|n|eB} + |\Delta|)^2 + m(n, k_z)^2}, & |n| \geq 1 \\ \pm\sqrt{v^2 k_z^2 + \Delta^2 + m(0, k_z)^2}, & |n| = 0. \end{cases} \qquad (7)$$

where $m(n,k_z) = m - Ck_z^2 - 2eBC(|n|+\frac{1}{2})$, and here we have assumed $v_x=v_y=v_z$ for simplicity. From topological insulators, we understand that the band inversion is the key to the realization of many exotic topological phases[45] and the nearly zero slope feature of the M1 transition indicates the transition energy does not change much with magnetic field which suggests the identical change of the initial and final Landau levels. Thus, it is of great significance to distinguish the inverted band structure from the normal band structure in experiments. The corresponding Landau fan diagrams are shown in Fig. 4b and 4c, where the zeroth Landau levels for the normal case and inverted case are quite different. For the normal case with quadratic dispersion (Fig. 4c), the energies of all Landau levels in particle (hole) branch increase (decrease) monotonically with the magnetic field and the slope ratio of transition $L_{-1} \to L_0$ and $L_{-2} \to L_1$ should be around 2. However, for the inverted case, the energy of the zeroth Landau level always shows an opposite monotonic behavior to other Landau levels belonging to the same branch. Consequently, the lowest spectrum corresponding to the optical transition $L_{-1} \to L_{0^+}$ or $L_{0^-} \to L_1$ can be very flat in a certain magnetic field regime leading to a large M2/M1 ratio. Based on the theoretical calculations, the spectra in Fig. 2a and 4a with finite zero-field intercepts can be understood. M1 and M2 are from a gapped band with band inversion. As shown by Fig. 4c and 4e and Supplementary Note XI, M4 and M5 are originated from normal (non-inverted) bands. The overall understanding of the Landau quantization from multiple bands are summarized in Fig. 5 and Table I.

Here we provide a general understanding of zeroth Landau levels in topological materials. In two-dimensional Dirac semimetals such as graphene or the surface state of three-dimensional topological insulators, the lower dimension makes Landau levels (including the zeroth one) non-dispersed along $k_z$ directions. Therefore, a gap will be generated between adjacent Landau levels. This effect is most prominent between the zeroth and the first Landau level. The transition energy thus follows $\sqrt{B}$. However, in conventional bulk materials, the zeroth Landau level is parabolically dispersed along $k_z$. Although the gap vanishes, density of state diverges at $k_z=0$, resulting in the resonant energy (between the zeroth and the first Landau level) following $B$. In the bulk state of topological insulator and the inverted gap state observed in NbAs, while the optical transition is still limited at $k_z=0$, the energy of the zeroth Landau level decreases with magnetic field due to the band inversion. In Weyl semimetal, the zeroth chiral Landau level is linearly dispersed in $k_z$, resulting in two distinct features: optical transition away from $k_z=0$; and the transition frequency and momentum position highly dependent on the Fermi level. The transition energy is a function of both magnetic field and Fermi level.

In conclusion, we have studied the quasiparticle dynamics with contributions from multiple bands in NbAs through magneto-optical spectroscopy. All the optical features are summarized in Fig. 2a, Fig. 5, Table I and Supplementary Note IV. Taking the advantage of probing Landau levels under high magnetic fields, unconventional optical transitions reveal the existence of the zeroth chiral Landau level. Different from other degenerated Landau levels with non-zero index, it unconventionally disperses linearly in $k_z$ direction and allows the optical transition without the limitation of $k_z=0$ or $\sqrt{B}$ evolution. Finite scattering helps to enhance the feature from chiral Landau level. Besides, two inequivalent Weyl nodes are found with different Fermi velocity. Fitting the magnetic field dependence of optical transitions unveils that the Weyl node in NbAs with higher Fermi velocity possesses much stronger particle-hole asymmetry. Massive Dirac fermions and massive trivial fermions are also found by observing the linear-in-$B$ cyclotron resonance. We summarize the overall Hamiltonian in Supplementary Note XIII and also provide a general picture of the zeroth Landau levels in topological materials. A comparison of the Landau quantization between Weyl fermions and massive fermions was made to further verify the unique feature of the zeroth chiral Landau levels.

**Method**
**Sample Growth.**
NbAs single crystals were grown by enhanced chemical vapor transport method using iodine as the agent with tilted ampules angle to maximize the growth rate and crystal quality. A temperature gradient is required to provide the necessary driving force for gaseous species in diffusion and connection between the cooler and hotter end. The stoichiometry and crystal structure were examined by XRD (Bruker D8 Discover) and EDX in a scanning electron microscopy.

**Magneto-optical measurements.**

The far-infrared and mid-infrared reflection were measured in a Faraday configuration with a superconducting magnet up to 17.5 T. NbAs samples were exposed to the globar infrared light through light pipes with (001) surface perpendicular to both the incident light and magnetic field. Infrared light was focused with parabolic cone. The reflected light was detected by a bolometer and analyzed by a Fourier transform infrared spectrometer (FTIR). All the light tube, samples and bolometer were kept at liquid helium temperature in a cryostat. The light path was pumped under vacuum to avoid the absorption of water and other gases.

**Zero-field reflection measurements.**

Zero-field reflection spectrum was measured with *E* // (001) surface at liquid helium temperature on a combination of Bruke 113V and 80V. An *in-situ* overcoating technique was used as reference.

**Magneto-transport measurements.**

Magneto-transport measurements were carried out using physical property measurement system and pulsed magnetic field up to 60 T at Wuhan National High Magnetic Field Center. A standard Hall-bar geometry is made. Lock-in amplifiers were used to measure the electrical signals.

**Data availability**

The data that support the findings of this study are available from the corresponding author upon reasonable request

**Acknowledgements**

This work was supported by the National Natural Science Foundation of China (Grant No. 11474058, 61674040, 11674189) and the National Key Research and Development Program of China (Grant No. 2017YFA0303302, 2016YFA0203900, 2017YFA0303504). We thank Ying Wang, Xue Jiang, Zhiqiang Li, Nanlin Wang for helpful discussions. H. Y. is also grateful to the financial support from Oriental Scholar Program from Shanghai Municipal Education Commission. A portion of this work was performed at the National High Magnetic Field Laboratory, which is supported by National Science Foundation Cooperative Agreement No. DMR-1157490, DMR-1644779 and the State of Florida. Part of the sample fabrication was performed at Fudan Nano-fabrication Laboratory.

**Author contributions**

F.X. conceived the idea and supervised the experiments.  Z. L. & X.C. carried out the growth. X.Y., C.Z., C.S., W.W., Y.L., J.L, D.S, T. X. and H.Y. performed the magneto-optical experiments. C.S. and H.Y. did magneto-Drude-Lorentz fitting. M.Z. carried out the zero field reflection measurements. X.Y, C.Z., Y.L, X.Z, Z.X. conducted magneto-transport experiments. Z.Y. and Z.W. performed theoretical calculations. X.Y., Z.Y., C.Z. and F.X. wrote the paper with helps from all other authors.

**Competing interests**

The authors declare no Competing Financial or Non-Financial Interests.

# FIGURES

**Figure 1 | Material characterizations and magneto-optical spectra. a**, Body-centered tetragonal structure of NbAs with inversion symmetry breaking. **b**, Single-crystal XRD spectrum, showing the (001) crystal planes of single crystallinity of NbAs. The inset is a photo of the sample with a length of 2 mm. **c**, Schematic plot of the magneto-optical experimental setup. **d**. Magneto-optical spectra of NbAs under different magnetic fields. The reflection rate under magnetic field versus photon energy is normalized by zero-field reflection measured from the same setup. The curves for different magnetic field are vertically shifted for clarity. These features with different magnetic field evolution are classified as C-type, M-type and T-type. Note that the dark arrows denote C3 transitions.

**Figure 2 | Inter-Landau-level transition energy from multiple carriers. a**, Transition energy versus $\sqrt{B}$ in the far-infrared range. Colored lines and labels are used to differentiate the origin of each feature. Red and blue dashed straight lines are guide lines for T-type transitions pointing to the origin of the plot, indicative of the two inequivalent massless Weyl fermions. Dark green dashed lines are fitting curves with linear-in-$B$ evolution for M-type transitions, suggesting gapped electronic bands with band inversion. Red and blue solid lines are fitting curves for the intra-band transition involving the zeroth chiral Landau levels in Weyl nodes. In this plot, colors in red, blue, dark green (for curve, guidelines and labels) denote the contributions from Weyl node 1 (W1), Weyl node 2 (W2) and gapped band, respectively. Inset shows that W1 has a higher Fermi velocity than W2. Labels in purple present that the feature is contributed by both Weyl nodes, thus mixing red and blue. The label $-n \to n+1$ denotes the initial and final Landau levels involved in the optical transition process. $-(n+1) \to n$ transition is also involved but not labelled for simplicity. Error bars are added based on both experimental resolution and fitting uncertainty. **b**, The normalized reflection data from experiments (black) and the Magneto-Lorentz fitting (red). The energy of the inter-Landau level transition in Fig. 2**a** is described by the fitted energy (arrows) instead of apparent peak maxima. 2% is the change of the reflection under magnetic field. **c,** Landau level dispersion of massless Weyl fermion in NbAs. Only the lowest nine branches are plotted for illustration. The arrows delineate the allowed optical inter-band transition processes between Landau levels at $k_z = 0$. **d, e,** Fan diagrams for W1 and W2 under different magnetic fields. Solid lines are guide lines pointing to the origin of the plot, indicative of the massless Weyl fermions and verifying the assignment of transition index. Labels indicate the transition and Landau level index.

**Figure 3 | Quantum limit and chiral Landau levels. a**, **b**, Longitudinal magneto-resistance with magnetic field up to 60 T and 10 T, respectively. The quantum limit is reached at 12 T and 19 T for two Weyl nodes. The inset shows the schematic drawing of the experimental setup. Arrows represent the peak position. The colors of the arrows denote the oscillation components from two inequivalent Weyl nodes. **c**, Landau fan diagram. Landau index is plotted against 1/B. Resistivity peak represents the integer Landau index. The observed intercept suggests non-trivial topological nature of both components in the SdH oscillations. **d,** Inter-Landau-level transition in terahertz range where the intra-band transition is allowed. Red and blue lines have the same meaning as Fig. 2a. Black dashed lines are guide lines (linear in $\sqrt{B}$), showing a deviation of chiral Landau level transitions from the normal ones. The fitting curve indicates the origin of C1/C3 and C2 from the zeroth chiral Landau level of W2 and W1, respectively. **e, f,** Schematic plots of Landau level dispersion for W1 and W2. The chiral Landau level is linearly dispersed. For W1 (not reaching the quantum limit), there exists two intra-band transition edges while W2, already in the quantum limit (*B*>12 T), has only one.

**Figure 4 | Landau fan diagram of massive fermions originated from inverted bands and normal bands according to Eq. (7). a,** Inter-Landau-level resonance energy versus magnetic field. For those transition already plotted in Fig.2a, error bars are not shown. Dashed straight lines with labels represent massive fermions with band gap. Color in dark green denotes the transition from the inverted gap structure. Color in grey describes the transitions from quadratic normal bands or undefined bands. The inset is the mid-infrared magneto-optic spectrum at 17.5T. **b**, **c,** the Landau fan diagrams of the inverted and normal case, respectively. **d, e**, the frequency versus $\sqrt{B}$ relation of the lowest two optical transitions in **b** and **c**, respectively. The parameters in Eq. (7) chosen for illustration are: for **b&d**, $Cm/v^2 = 0.5$, $\Delta/m = 0.5$; for **c&e**, $Cm/v^2 = -0.5$, $\Delta/m = 0.5$; $k_0 = \sqrt{m/2C}$. Note that the qualitative difference between the inverted case and normal case does not depend on the parameters chosen.

**Table I. Landau quantization of the quasi-particles summarized in NbAs with different topological nature.** The energies of the Landau level are plotted versus $k_z$. Dashed lines are Fermi level. Arrows represent the allowed optical transitions. Red and blue colors denote the transition in Weyl node 1 and Weyl node 2, respectively. Green and grey colors represent the transitions in gapped electronic band with and without band inversion, respectively.

| Optical Transition | T1 | C1-C3 | T2-T7 | M1-M2 | M4-M5 |
|---|---|---|---|---|---|
| Band | Weyl1 | Weyl1 & Weyl2 | Weyl1 & Weyl2 | Inverted Gap | Topological Trivial Gap |
| Landau level Dispersion | Normal | Linear & Chiral | Normal | Normal | Normal |
| Transition | Intra-band | Intra-band | Inter-band | Inter-band | Inter-band |
| $B$ evolution | $\sqrt{B}$ | $\pm\sqrt{c^2 \pm B} \mp c$ | $\sqrt{B}$ | $\sim c$ | $B$ |

*B* and *c* denote the magnetic field and arbitrary constant, respectively. Note that for simplicity the equation for C1-C3 does not include particle-hole asymmetry.

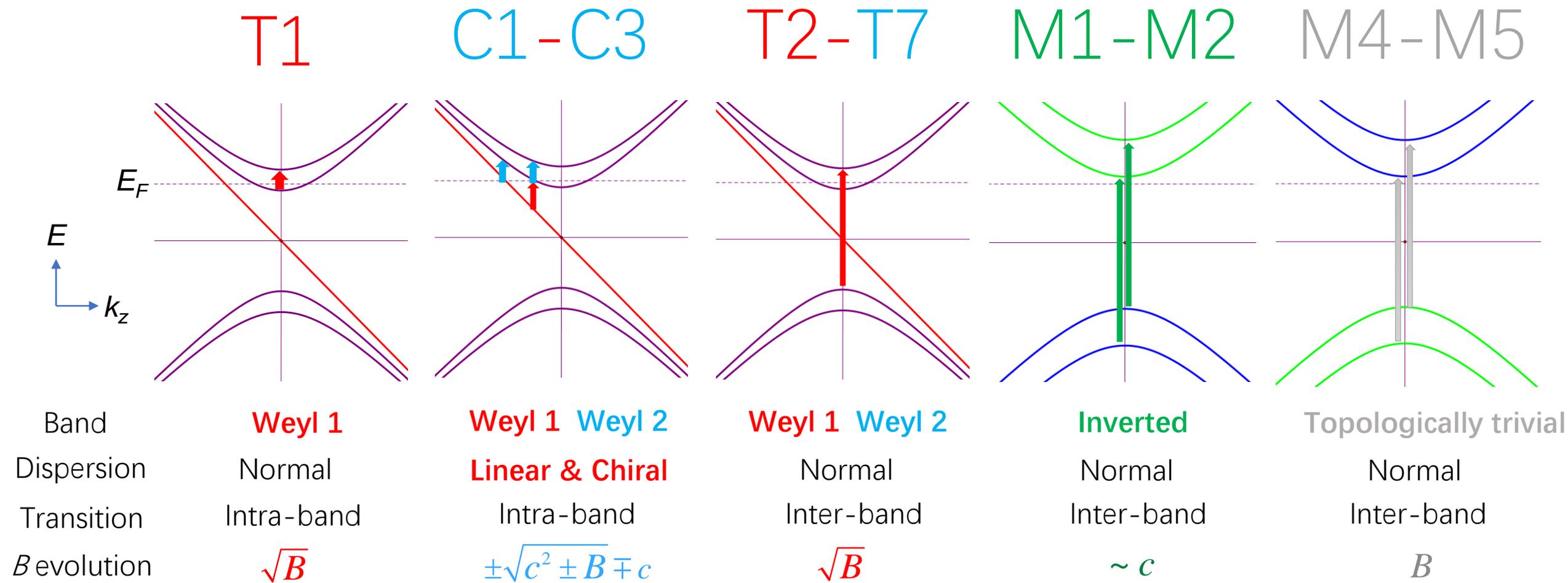

Table I

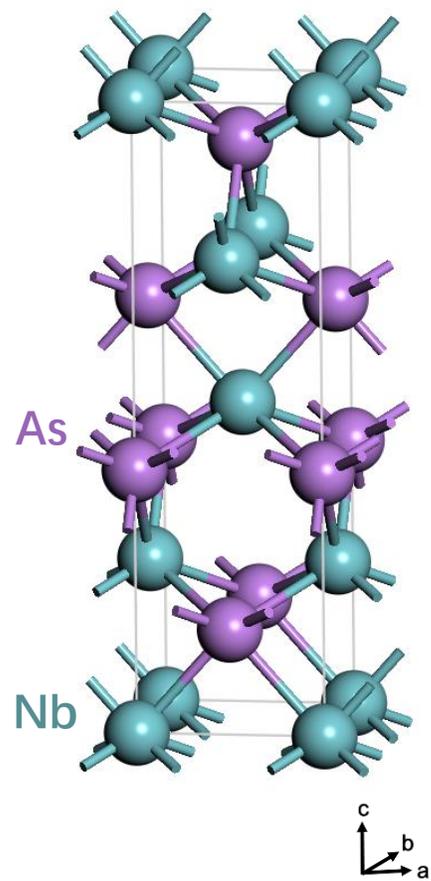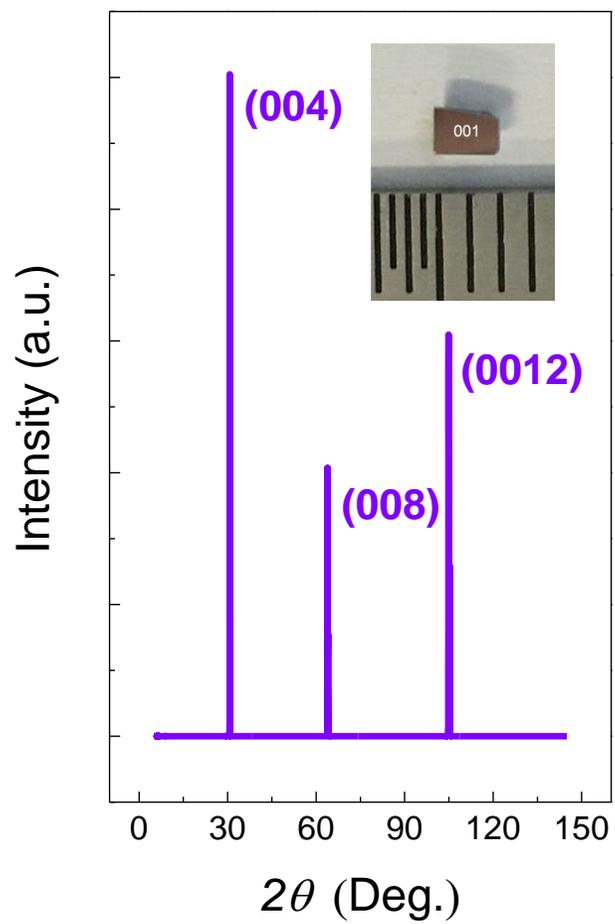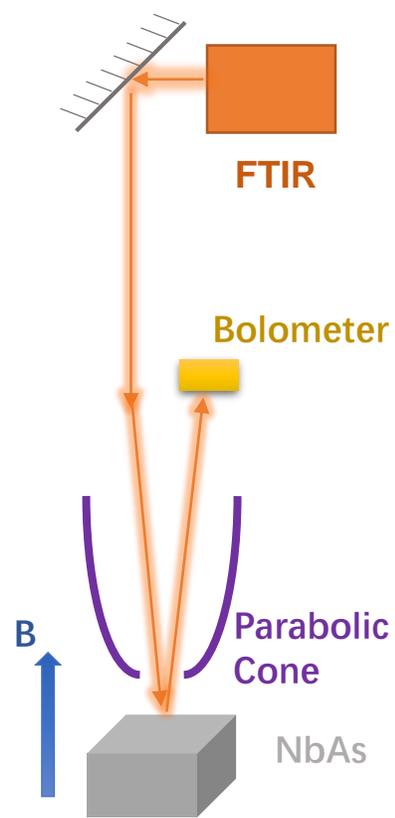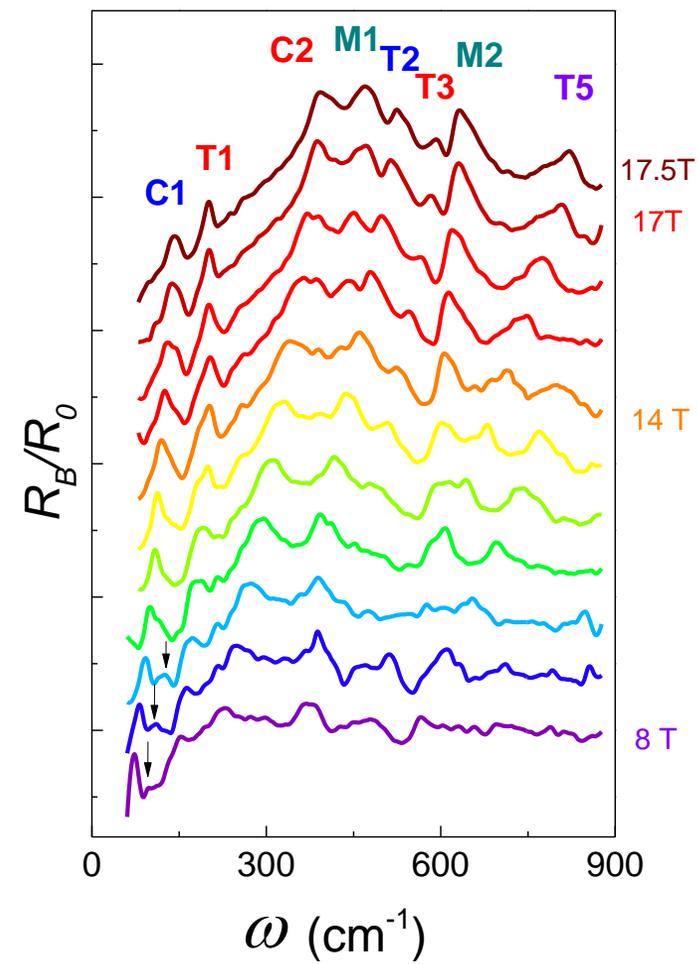

Figure 1

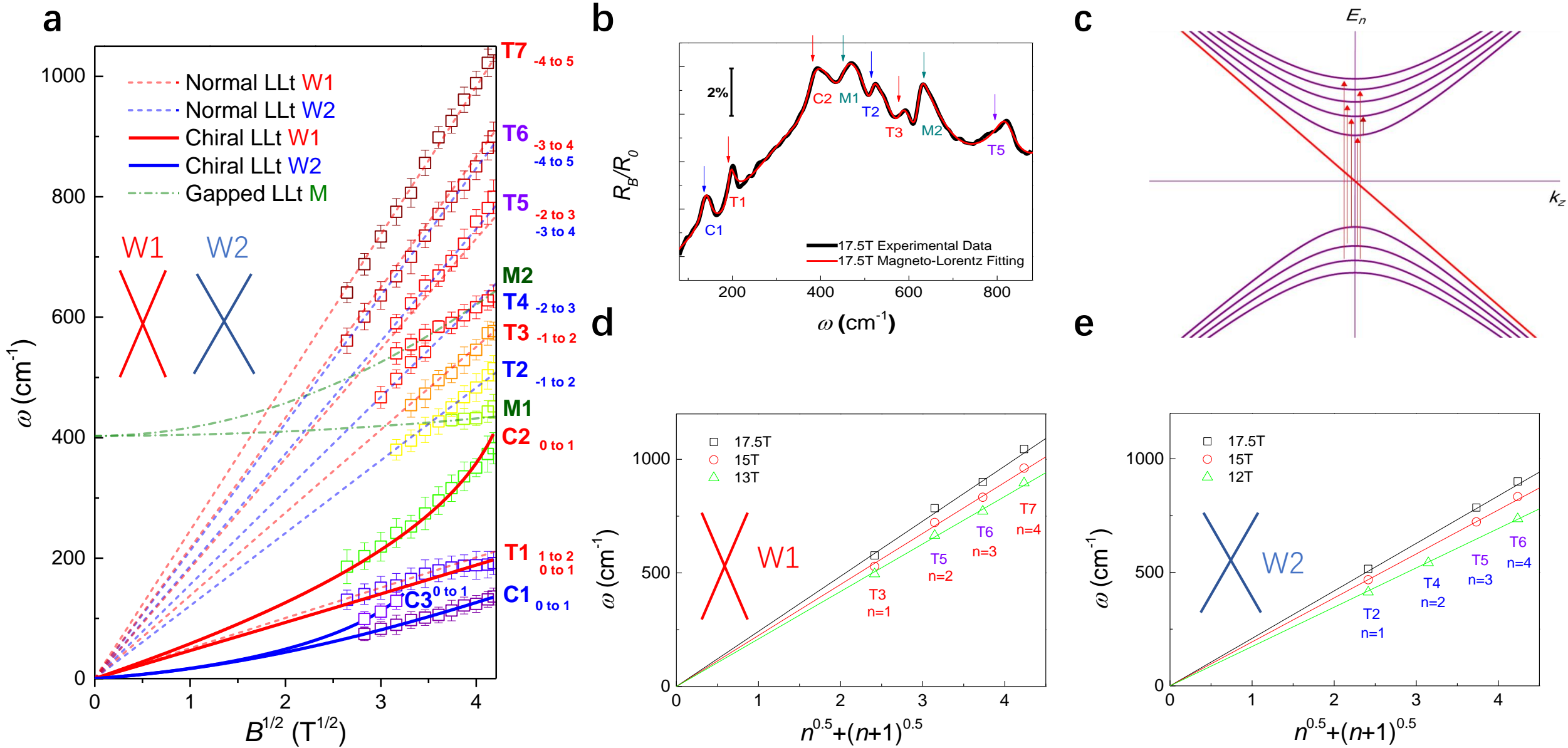

Figure 2

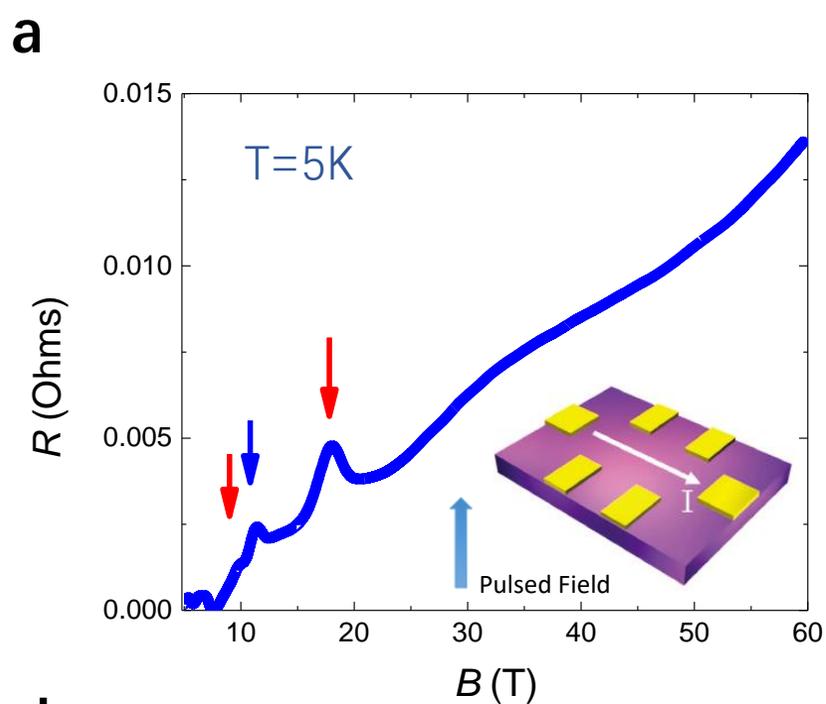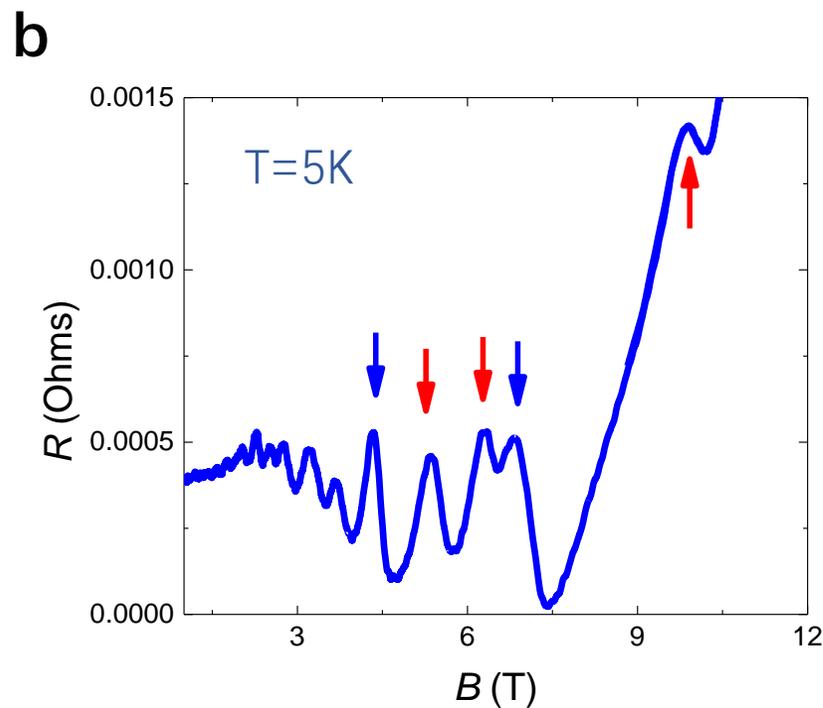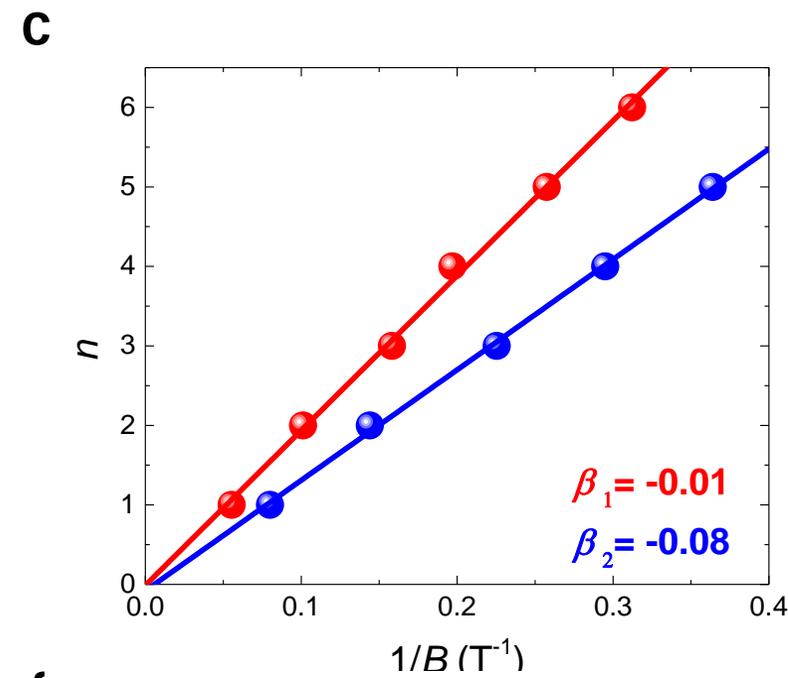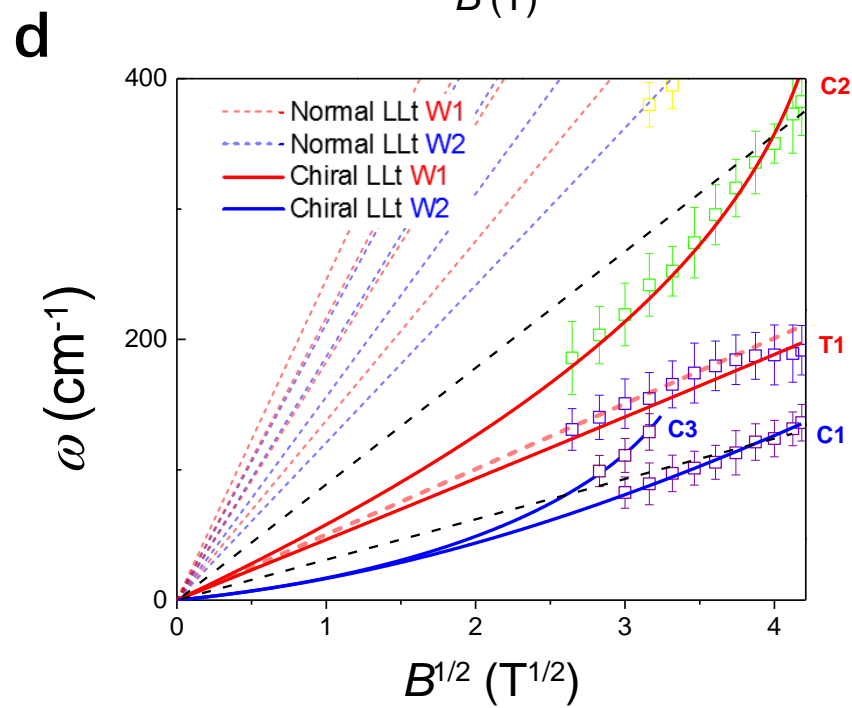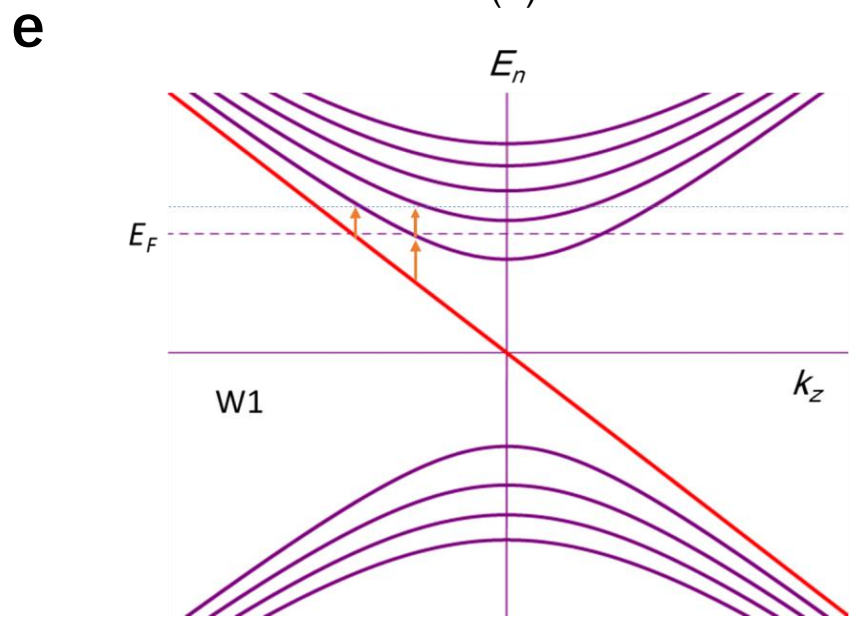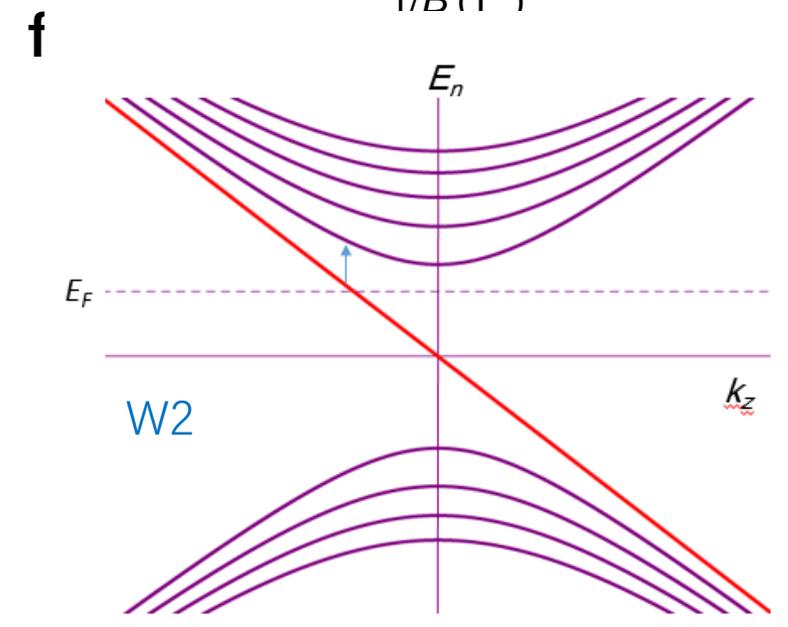

Figure 3

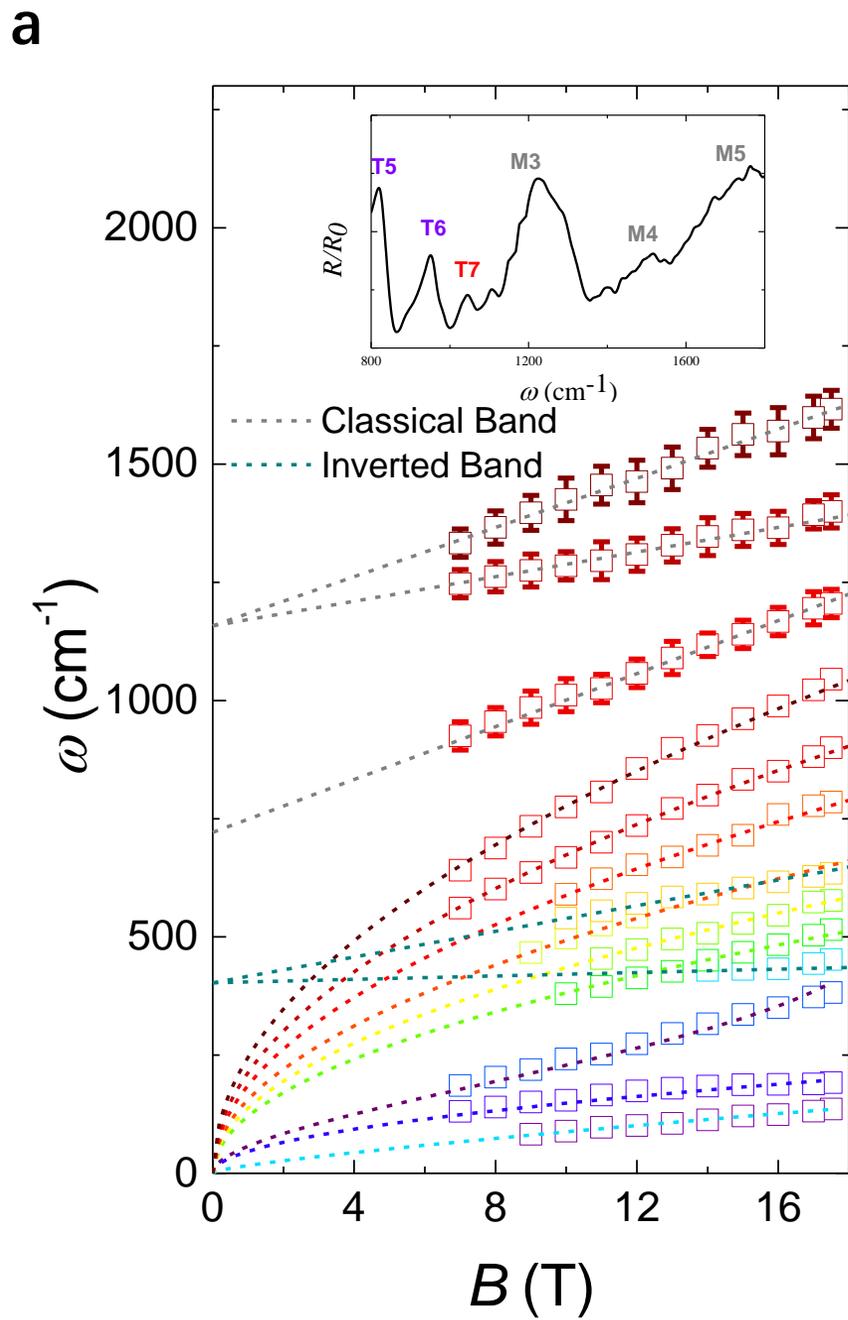
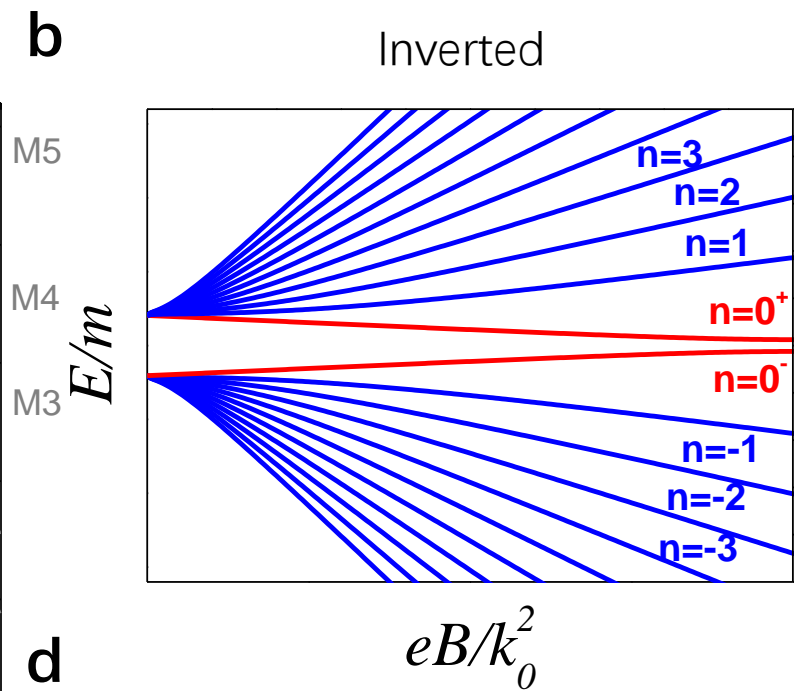
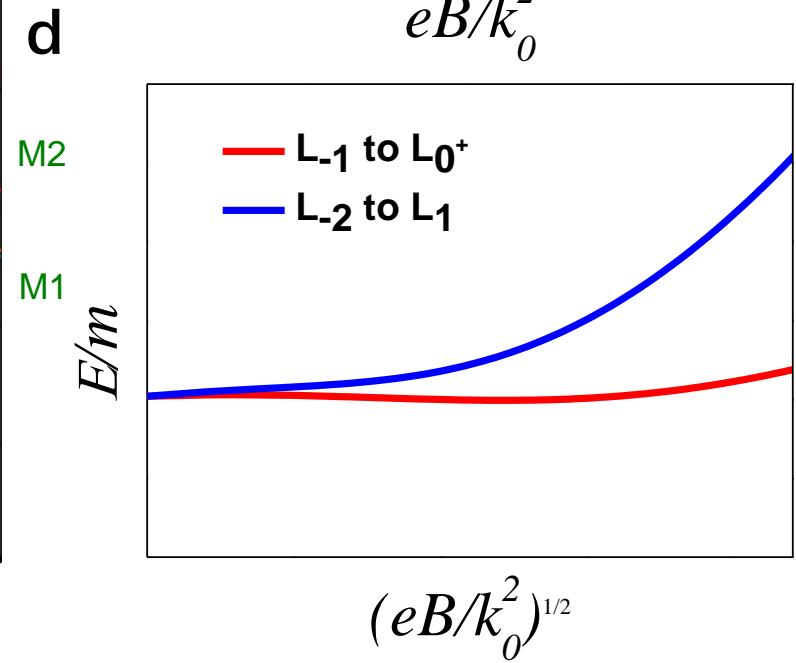
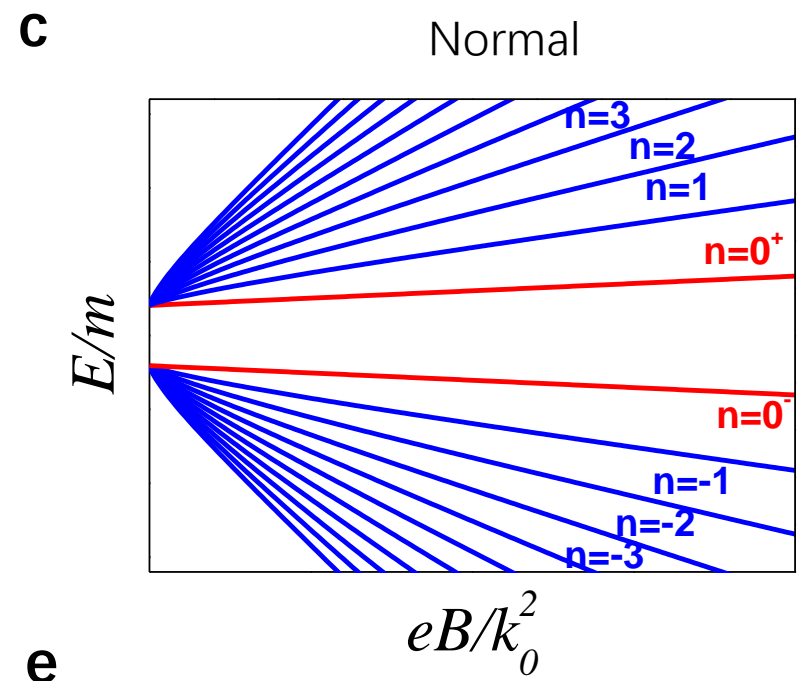
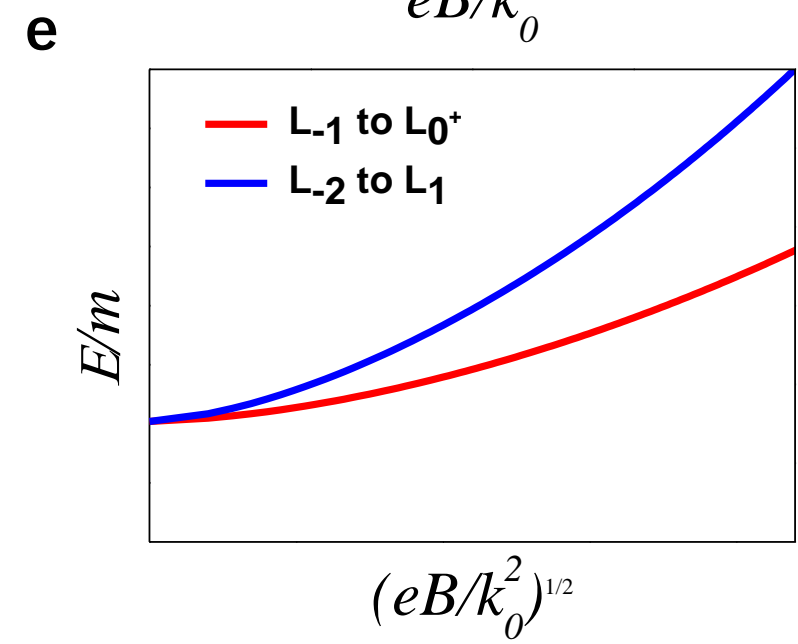

Figure 4